\input epsf
\documentstyle{l-aa}

\newcommand{\beq}{\begin{equation}}
\newcommand{\eeq}{\end{equation}}
\newcommand{\eeql}[1]{\label{eq:#1}\end{equation}}

\newcounter{compteur}

\def\mv{M_V}
\def\mk{M_K}

\def\814{I_{814}}
\def\606{R_{606}}
\def\555{V_{555}}
\def\mh{[M/H]}

\def\msol{\mbox{M}_\odot}

\def\te{T_{\rm eff}}
\def\simgr{\,\hbox{\hbox{$ > $}\kern -0.8em \lower 1.0ex\hbox{$\sim$}}\,}
\def\simle{\,\hbox{\hbox{$ < $}\kern -0.8em \lower 1.0ex\hbox{$\sim$}}\,}
\def\wig#1{\mathrel{\hbox{\hbox to 0pt{%
          \lower.5ex\hbox{$\sim$}\hss}\raise.4ex\hbox{$#1$}}}}

\def\aj{AJ}                  
\def\araa{ARA\&A}             
\def\apj{ApJ}                 
\def\apjl{ApJ}                
\def\apjs{ApJS}               
\def\aap{A\&A}                
\def\aaps{A\&AS}             
\def\mnras{MNRAS}             
\def\pasp{PASP}               

\begin{document}
\thesaurus{}
\title{ 
Evolutionary models for solar metallicity low-mass stars: mass-magnitude relationships and color-magnitude diagrams}

\author{{\sc I. Baraffe\inst{1}, G. Chabrier\inst{1}, F. Allard\inst{1,2} and
 P. H. Hauschildt\inst{3} }} 

\institute{
C.R.A.L. (UMR 5574 CNRS),
Ecole Normale Sup\'erieure, 69364 Lyon Cedex 07, France,
(ibaraffe, chabrier, fallard @ens-lyon.fr) \and
Dept. of Physics, Wichita State University,
Wichita, KS 67260-0032,
(allard@eureka.physics.twsu.edu) \and Dept. of Physics and Astronomy
and Center for Simulational Physics, University of Georgia
Athens, GA 30602-2451,
(yeti@hobbes.physast.uga.edu)}

\date{Received date ; accepted date}

\maketitle

\markboth{I. Baraffe et al.: Evolutionary models for solar metallicity low-mass stars } {}

\begin{abstract}

\bigskip
We present evolutionary models for low mass stars from 0.075 to 1 $\msol$
 for solar-type metallicities [M/H]= 0 and -0.5. 
The calculations include the most recent interior physics and the latest generation of 
{\it non-grey} atmosphere models. We provide mass-age-color-magnitude relationships 
for both metallicities. 
The mass-M$_V$ and mass-M$_K$ relations are in excellent agreement with the empirical relations derived observationally. 
The theoretical color-magnitude diagrams are compared with the sequences of globular clusters (47 Tucanae) and open
clusters (NGC2420 and NGC2477)
observed with the  Hubble Space Telescope. 
Comparison is also made with field star sequences in $\mv$-$(V-I)$,
$\mk$-$(I-K)$ and
$\mk$-$(J-K)$ diagrams. 
These comparisons show that the most recent improvements performed in low-mass star atmosphere models 
yield now reliable stellar models in the near-infrared. These models can be used for metallicity, mass, temperature and luminosity calibrations. 
Uncertainties still remain, however, in the optical spectral
 region below $\te \sim
3700K$, where predicted (V-I) colors
are too blue by 0.5 mag for a given magnitude. The possible origins for such a discrepancy, most likely 
a missing source of opacity in the optical and
the onset of grain formation are examined in detail.

\bigskip

\keywords{stars: Low-mass --- stars: evolution
--- stars: colour - magnitude diagrams --- stars: mass - magnitude relationship}
\end{abstract}

\section{Introduction}

The numerous data obtained within the past few years with ground-based and 
space-based near infrared projects provide nowadays a wealth of 
low-mass star observations from 1 $\msol$ down to the brown dwarf regime.
Observations cover a wide range of stellar populations, belonging
to young, open or globular clusters and to halo
and disk fields. Their analysis requires accurate
theoretical models spanning a large range of ages, masses and metallicities. 
Important progress has been realized recently on the theoretical side, which
emphasizes the complex physics involved in the modeling of these cool and dense objects. Recent work has demonstrated the
necessity to use accurate
internal physics and outer boundary conditions based on non-grey 
atmosphere models to describe correctly the mechanical and thermal properties of low mass objects (Burrows et al. 1993; Baraffe et al. 1995, 1997; Chabrier and Baraffe 1997, CB97). The tremendous efforts accomplished
recently in the modeling of atmosphere models and the
derivation of synthetic spectra (see the review by Allard et al. 1997), combined with interior models, now provide synthetic colors and magnitudes which can be compared directly to observed quantities, avoiding
the use of uncertain empirical $\te$ and bolometric
correction scales.

In a recent paper (Baraffe et al. 1997, BCAH97), we have derived evolutionary models for
metal-poor low mass stars based on the stellar interior physics
described in CB97 and on the Allard and Hauschildt (1998) "NextGen"
atmosphere models. Comparison with the lower Main Sequence of
 globular clusters observed with the HST has assessed
the validity of the models in the metallicity-range $-2.0\le [M/H]\le -1.0$.
The success of these models has been confirmed recently by new observations of NGC6397 down to $\sim 0.1 \,\msol$ (King et al., 1998), but more importantly with the observations realized with the NICMOS camera. Indeed, at the time of the BCAH97 analysis, only optical (V-I) colours were available for the clusters.
Recent observations performed with NICMOS for the first time
provide colour-magnitude diagrams (CMDs) in the near-infrared domain for
$\omega$Cen (Pulone et al. 1998) and NGC6397 (Paresce, priv. com.). The agreement with the models 
is excellent and the observations confirm in particular the predicted blueshift in IR colors near
the bottom of the main sequence, which stems
from 
ongoing collision-induced  absorption of molecular hydrogen (see BCAH97 Fig. 7). This, we believe, assesses the reliability of our
metal-poor models down to the bottom of the main sequence.

The natural continuation of this work is the extension to solar-like metallicities.
This is the aim of the present paper. The present calculations are based on the same microphysics, described in CB97, and are confronted to available observations in the range $-0.5\le [M/H]\le 0$.
The calculation of atmosphere models for solar metal-composition is rendered more complex by the importance of molecular metal-bands, which shape the emergent spectrum.  
In this range of
metallicity, the stellar spectra and atmospheric structures become
very sensitive to the treatment of molecular opacity,
dominated by H$_2$O in the IR and TiO and, to a less extent, VO in the optical. It is thus essential to confront theory with observations at these wavelengths to 
determine the remaining uncertainties in the models for solar-metallicity. 
We first summarize the physics entering specifically the solar models
(\S2). In \S 3, we compare the theoretical mass-magnitude relationships 
in the V- and K-bands with the observationally-derived relationships.
In \S 4, we
compare the results with
observed CMDs of (i) the globular cluster 47 Tucanae with [M/H] $\sim$ -0.5, (ii) two open clusters observed with the HST, 
$NGC2477$ and $NGC2420$ with [M/H] $\sim$ 0, 
and (iii) disk field stars in optical and near-infrared colors.
Section 5 is devoted to the conclusion.

\section{Theory}

A complete description of the physics involved in the models is given in CB97 and the atmosphere models are described in
Allard and Hauschildt (1998, AH98). We only briefly summarize the main inputs. 
These models are based on
the most recent physics characteristic of low-mass star interiors,
equation of state (Saumon, Chabrier \& VanHorn 1995, SCVH EOS), enhancement
factors of the nuclear rates (Chabrier 1998) and updated opacities
(Iglesias \& Rogers 1996; OPAL), the last generation of non-grey
atmosphere models (Allard et al. 1997; AH98) and accurate
boundary conditions between the interior and the atmosphere
profiles (CB97; BCAH97).

As shown in detail in \S 2 of CB97, the pure hydrogen-helium SCVH EOS remains valid to describe the interior structure and the evolution of objects with solar metal-abundances,
given the very small {\it number}-abundance of metals.
Comparison with an EOS including metals
(e.g. Mihalas, D\"appen \& Hummer 1988; MHD) in its domain of validity ($\simgr$ 0.4 $\msol$) reveals
differences of less than 1.3\% in $\te$ and 1\% in $L$ for a given
mass. 

CB97 have also examined extensively the effect of the outer boundary conditions
between the interior and the atmosphere profiles (see their \S 2.5). These
conditions are crucial since they determine
the mass-$\te$ and $T_{int}$-$\te$ relationships. These authors have demonstrated that {\it any} grey approximation
is physically incorrect below $\te \sim 5000 - 4500$ K
($\sim 0.6 - 0.8\,\msol$), depending on the metallicity, (see also
Fig. 1 below)
and have stressed the {\it necessity} to use non-grey atmosphere models and accurate
boundary conditions
to derive reliable low-mass star models.
As shown by these authors, a grey approximation yields denser and cooler
atmosphere profiles below the photosphere and thus {\it overestimate the effective temperature,
and the luminosity} for a given mass (see Fig. 5 of CB97), yielding
erroneous evolutionary tracks and mass-luminosity relationships.
This is clearly illustrated in Figure 1 which compares the effective temperature
vs central temperature relationship in
the low-mass star regime down to the brown dwarf limit when using either consistent
non-grey boundary conditions or a grey approximation. 
The models are displayed in Fig. 1 for an age t=5 Gyrs.
We note several modulations
in these relations which stem from the very atmospheric and/or internal properties
of these stars. 
The change of slope below 0.8 $\msol$ reflects the decreasing
efficiency of hydrogen burning at t=5 Gyrs, leading to a steeper drop
of $\te$ with $T_c$. When H burning is efficient,  a decrease of the central hydrogen abundance X$_H$
in the radiative core, and thus an increase of the molecular weight $\mu$
results in an increase of $T_c$ and thus L ($L\propto T_c^4 \propto \mu^4$
 for radiation),
which provokes the expansion of the envelope. This increase of the
total radius of the star leads to a less steep increase of $\te$
($\te \propto L^{1/4}/R^{1/2}$). 
Conversely, {\it below} 0.8 $\msol$, the increasing central
abundance of hydrogen with decreasing mass leads to the faster drop of $\te$ with $T_c$ displayed
in Fig. 1.
The change of slope 
around $\sim 0.5 \,\msol$ reflects the onset of H$_2$ molecular formation near
the photosphere. This lowers the adiabatic gradient and
favors the onset of convection in the atmosphere
(Auman,
1969; Copeland, Jensen \& J\o rgensen, 1970; Kroupa, Tout \& Gilmore, 1990).
The central temperature keeps
decreasing but, given the efficiency of convective transport, the effective temperature does
not drop as quickly as it would without H$_2$.
The last drop below $\sim 0.2\,\msol$ reflects the overwhelming
importance of electron degeneracy in the interior (see CB97).

\begin{figure}
\epsfxsize=95mm
\epsfysize=95mm
\epsfbox{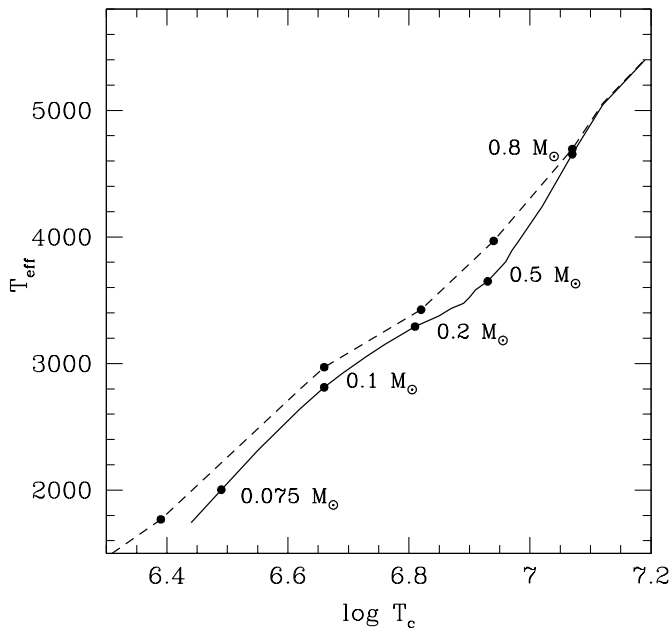} 
\caption[ ]{ Central temperature-effective temperature relation for $\mh$=0
 for stellar models based on different outer boundary conditions: consistent
boundary conditions (see CB97 \S 2.5)
with the present non-grey atmosphere models
(solid line, this paper); grey (Eddington)
approximation (dashed line). The changes of slope are discussed in
the text.
}
\end{figure}

The "NextGen" atmosphere models of Allard and Hauschildt (1997) represent a substantial improvement with respect to the previous
so-called "Base" generation models (Allard and Hauschildt, 1995) and  cover a wide range of temperatures and metallicities. The treatment of pressure broadening and of molecular line absorption coefficients is considerably improved in the "NextGen"
models and leads to physically more reliable atmosphere models
(see Allard et al. 1997, for details). 
A comparison of the mass-$\te$ relationship obtained with stellar models
based on both sets of atmosphere models is shown in Figure 2. A preliminary set of the present improved stellar models has been presented in Chabrier, Baraffe
\& Plez (1996, CBP96) and has been shown to yield a mass-M$_V$ relationship in better agreement with the observationally-derived relation of Henry and Mc Carthy (1993, HMC93) than models based on the "Base" atmospheres.
 The main difference between this preliminary set of the "NextGen" atmosphere models and the present one is the different water line list, the Jorgensen (1994) one
in CBP96 and the Miller et al. (1994) one in the present models. As shown in Fig. 2,
evolutionary models based on both sets of atmosphere models are very similar
in a global, bolometric diagram,
but the Miller et al. (1994) water linelist yields a substantial improvement in the K-band, illustrated by the good match with the empirical mass-M$_K$ relationship, as will be shown in \S3. CBP96 also presented solar metallicity models based on non-grey atmosphere models computed by Brett and Plez (Plez et al. 1992; Brett, 1995; Plez, 1995, priv. comm., BP95).
We recall that the BP95 and the "NextGen" models are based on different TiO and H$_2$O line lists (see CBP96 for details). Figure 2 shows that evolutionary models based
on the BP95 atmosphere models lie between the "Base" and the "NextGen"
models. The different sets of models and corresponding synthetic colors will be examined in \S 3 and \S4. It is important at this stage to stress an essential point
: since the atmosphere profiles fix the outer boundary condition for the stellar interior
and are used to compute the synthetic spectra, it is {\it essential}, for
sake of consistency when deriving theoretical mass-magnitude-color relationships, to adopt
the boundary conditions and the synthetic colors based on the {\it same atmosphere structures}. Any model providing mass-$\te$ relationships based on a given set of atmosphere models and using
synthetic colors based on a different set to derive the $\te$-color relationship
is severely inconsistent and thus inaccurate.

\begin{figure}
\epsfxsize=95mm
\epsfysize=95mm
\epsfbox{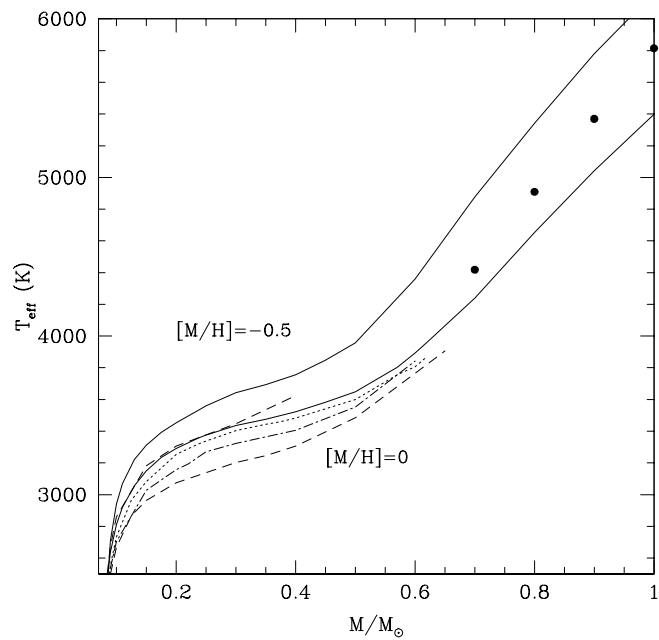} 
\caption[ ]{ Mass-effective temperature relation for $\mh$=-0.5
(top curve) and $\mh$=0 (bottom curve) for stellar models based on different atmosphere models: "NextGen" (solid line, this paper);
"Base" (dashed line, BCAH95); preliminary set of the "NextGen"
models with the Jorgensen (1994) water linelist for [M/H]=0 (CBP96; dotted line); BP95 for [M/H]=0 (dash-dotted line). The full circles correspond to the present models with inputs specific to the Sun
(i.e $\alpha=1.9$, Y=0.282, see text).

}
\end{figure}

We have computed grids of models from 1 $\msol$ down to the brown dwarf limit,
i.e $\sim 0.075 \msol$ for [M/H]=0 and  $\sim 0.079 \msol$
for [M/H]=-0.5. The theoretical characteristics of the present models,
effective temperature, gravity, bolometric magnitude and magnitudes in VRIJHK
 for several ages are listed in Tables 1-2.
The "NextGen" models have been extended to higher effective temperatures,
enabling us to extend the mass grid up to 1 $\msol$, instead of
0.8 $\msol$ in CB97. Note that the 0.8 $\msol$ with [M/H]=0 in CB97 (cf their Table.2) is calculated
with an outer boundary condition based on a T($\tau$) relationship. For solar metallicity the results 
for this range of $\te$ ($\te > 4500K$)  
are similar to models based on non-grey 
atmosphere models (see Fig. 1), since (i) atmospheric convection remains
 below the photosphere and (ii) molecules, which introduce strong
non-grey effects, are stable only in the outermost layers. We stress,
however, that this limit depends on the metallicity. For
 sub-solar abundances, this occurs for {\it higher} temperatures, $\te \simgr $ 5000 K (cf CB97).

We use a helium abundance Y=0.275 (resp. 0.25) for [M/H]=0 (resp. -0.5) and
a general mixing length parameter $\alpha = l/H_p$ = 1. As shown in CB97 and BCAH97, although
variations of this parameter around standard values (within $\sim$ a factor 2)
are inconsequential below $\sim 0.6\,\msol$, they become important above this limit, as examined below.
We thus have also computed models from 0.7 to 1 $\msol$ based on the
inputs which reproduce the properties of the Sun at 4.61 Gyrs, namely
 $\alpha$ = 1.9 and Y= 0.282. Such models are given in Table 3 and shown in Figure 2 (filled circles).
Note that when using the MHD EOS, the
solar model
is fitted with $\alpha$ = 2 and Y =0.28 with an outer boundary
condition based
on detailed atmosphere models, and with $\alpha$ = 1.6 and Y =0.28 when using
 a T($\tau$) relationship (e.g. the Eddington approximation).
As seen from a comparison between Tables 1 and 3, the variation of the
mixing length parameter from 1 to 1.9 translates into a 9\%
difference in L and 4\% to 7\% in $\te$ above $ 0.7\, \msol$.

\section{The mass-magnitude relationships}

In this section, we analyse the theoretical mass - magnitude relationships and compare them to the observational data of Henry and Mc Carthy (1993, HMC93).
CPB96 have already presented the theoretical mass-$\mv$ relationship
and its comparison with the empirical fit derived by HMC93.
As stressed by these authors, the HMC93 fit is just an average
fit among the data and must be considered only as a (useful) guideline, for it
does not take into account physical differences due to age and
metallicity. Although the observational data suffer from large uncertainties 
 on the mass determination, they still
provide stringent constraints for the structure and the evolution of low-mass star models, in particular in the crucial region near the bottom of the main sequence. As mentioned in CBP96, although the agreement between
theory and observation at this time was excellent in the V-band, the
theoretical relationship was about 0.5 mag fainter than the empirical
one below $\sim 0.5\, \msol$ in the K-band. It clearly reflected the overestimated water absorption in these model atmospheres.
The inclusion of the Miller et al. (1994)
water line list in the present atmosphere models clearly solves this shortcoming
and yields
now an excellent agreement with the data set and the empirical relation derived
by HMC93. Both m-M$_V$ and m-M$_K$
relations are displayed in Figures 3a,b. Different ages are displayed
for the present  solar metallicity  models.

A striking feature illustrated in Figures 3a,b is the very weak
metallicity-dependence in the K-band, which becomes unobservable below
$\sim 0.4 \msol$ ($\te \sim$ 3500 K), compared to the strong dependence in the
V-band.
This comes from the fact that
below $\te \sim 3500 K$, which corresponds to the onset of molecular formation,  the opacity in the V-band, dominated by TiO and VO, increases with metallicity so that
 the peak of the energy distribution is shifted toward larger wavelengths, in particular to the K-band (see e.g. Fig. 3 of Allard et al., 1997).
This yields a decreasing V-flux and an increasing K-flux with increasing metallicity. On the other hand, for a given mass, the effective temperature decreases with increasing metallicity (see e.g. CB97 Fig. 13 and Tables 1-2) so that
the total flux decreases (F $\propto \te^4$). These two effects compensate in the K-band, yielding similar K-fluxes for $\mh=-0.5$ and 0, whereas they add up in the V-band, resulting in the important signature of metallicity in this passband. Note that the previous arguments remain valid as long as
H2 collision-induced absorption does not depress significantly the flux in the K-band, 
which is the case for metal poor low mass stars at the bottom of the Main Sequence. 

\begin{figure}
\epsfxsize=95mm
\epsfysize=95mm
\epsfbox{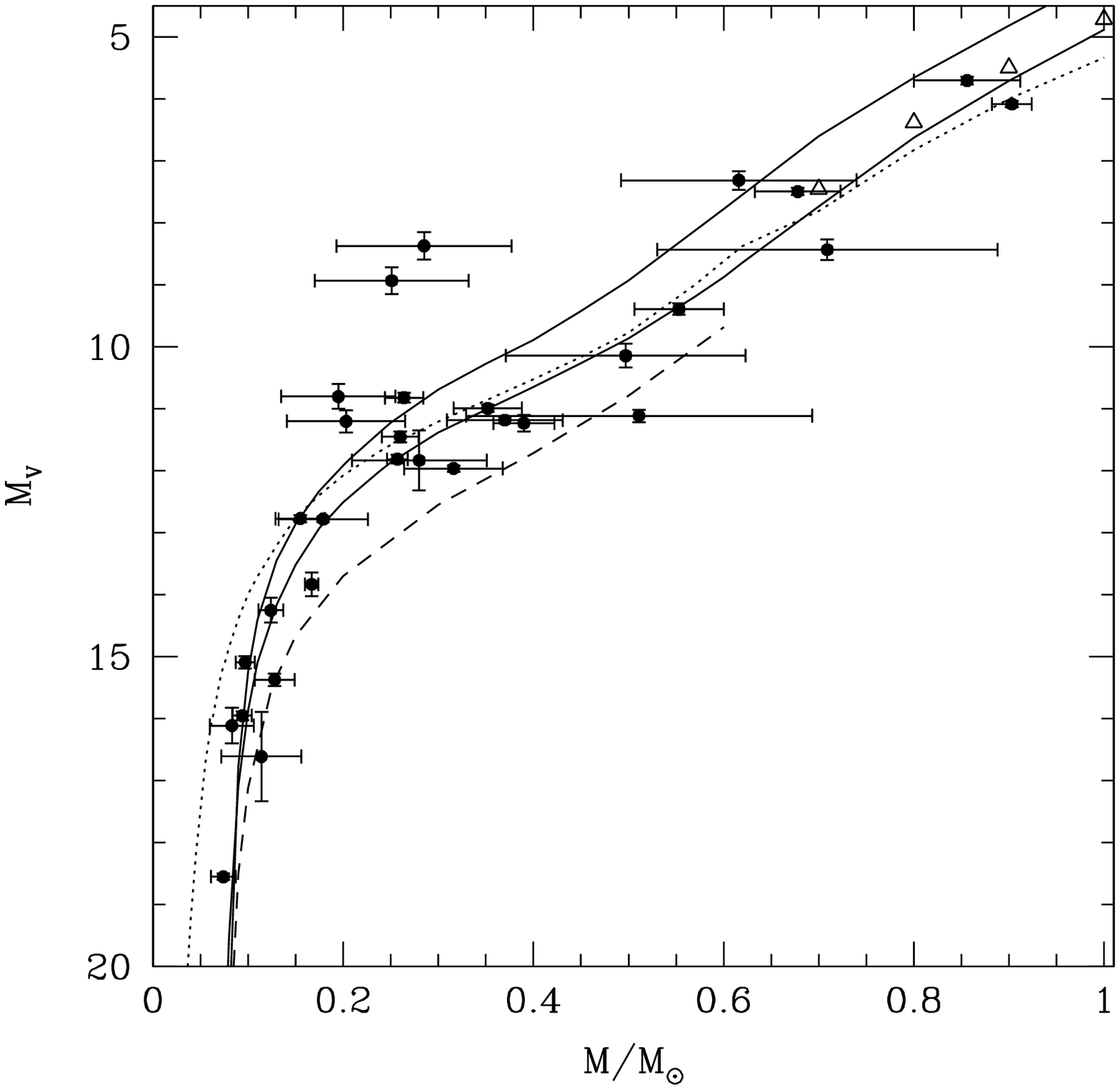} 
\epsfxsize=95mm
\epsfysize=95mm
\epsfbox{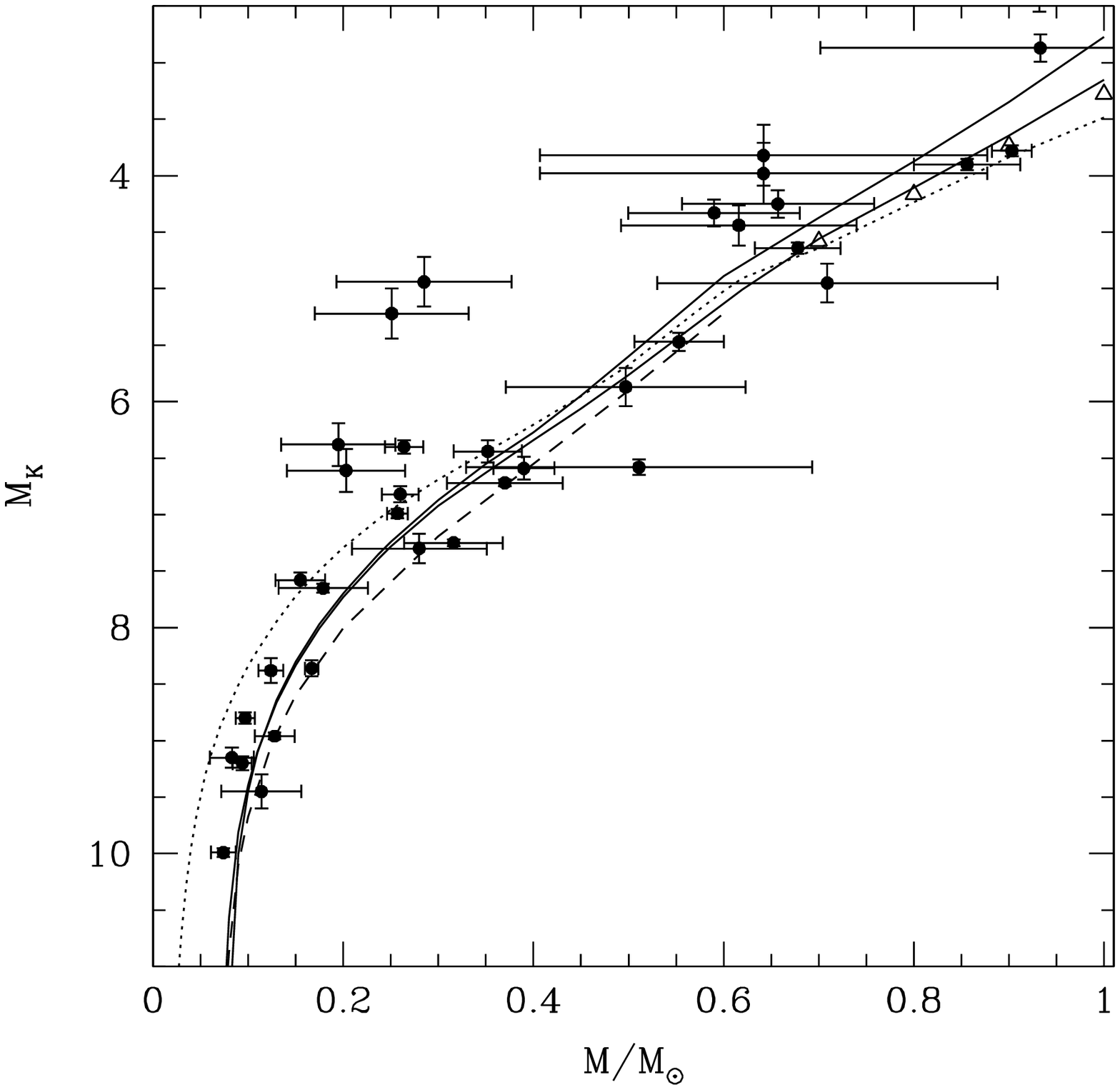}
\caption[ ]{ {\bf (a)} Mass-$\mv$ relation. The full circles are the data of
 Henry and Mc Carthy (1993). Solid lines: present models for [M/H]=0 (lower line) and [M/H]=-0.5 (upper line). Dashed line:
"Base" models for [M/H]=0. The open triangles correspond to present solar metallicity models with inputs specific to the Sun ($\alpha=1.9$, Y=0.282, see text).
The  models above  correspond to t=5 Gyrs. Dotted line: present [M/H]=0
models for t=100 Myrs. We recall that the zero-age main-sequence for
a 0.075 $\msol$ star is $\sim$ 3 Gyr. {\bf (b)} Same as Fig. 3a for the
Mass-$\mk$ relation.
}
\end{figure}

\section{Color-magnitude diagrams} 

\subsection{Open and globular cluster main sequences}

{$\bullet $} {\it 47 Tuc} (Figure 4) : This globular cluster has been observed  recently by Santiago et al. (1996) with the HST {\it Wide Field and Planetary
Camera-2 (WFPC2)} in the $F606W$ and
$F814W$ filters. Thanks to the courtesy of G. Gilmore and R. Elson, we are able to
compare the models with observations in
these bands, in the so-called WFPC2 {\it Flight}
system, as done for metal-poor clusters in BCAH97.  We use the analytical relationships of Cardelli et
al. (1989) to calculate the extinctions from the M-dwarf synthetic
spectra of Allard \& Hauschildt (1998) over the whole frequency-range,
with the reddening value $E(B-V)$ quoted by the observers.
The observed data, corrected for reddening, are then compared with the models.
We adopt the distance modulus (m-M)$_0$ = 13.38 and reddening E(B-V)=0.04 of Santiago et al. (1996). This yields the following extinction corrections: A$_{555}$ ($\sim$ A$_V$) = 0.125,  
 A$_{606}$ = 0.115 and A$_{814}$ ($\sim$ A$_I$) = 0.08, where V and I refer
to the standard Johnson-Cousins filters. The iron abundance for the cluster is [Fe/H] $\sim -0.7, -0.65$ (Santiago et al. 1996; Caretta \& Gratton 1997), which corresponds to a total metallicity [M/H] $\sim -0.5$ when taking into account the overabundance of $\alpha$-elements (Ryan \& Norris, 1991; see BCAH97). 

Figure 4 displays the results for t=10 and t=14 Gyrs. The age effect
is negligible below the turn-off mass m$_{TO}$
$\sim 0.85-0.9 \msol$ at t=10 Gyrs and $\sim 0.8 \msol$ at t=14 Gyrs.
Comparison is also made with models calculated with a mixing length
parameter $\alpha$ = 2, which affects the evolution only for $m>$ 0.6 $\msol$. As shown in the
figure, the theoretical isochrones calculated with $\alpha=1$ can be up to $\sim$ 0.05 mag too red in the uppermost main sequence ($M_{814} \simle 6.5$ i.e $m\simgr 0.6 \msol$) and up to $\sim$ 0.1 mag too blue in the lower main sequence ($M_{814} \simgr 7.5$). Adopting a 0.2 mag larger distance modulus 
would bring the upper main sequence in good agreement, but the lower
main sequence would be 0.08 mag too blue. A possible
calibration problem of the data in the HST filters seems unlikely since
we have used observations from the same group, with the same filters and calibration transformations, for $\omega$-Cen  ([M/H] $\sim$ -1),
and agreement between theory and observations is excellent (cf. BCAH97). Although the differences in the upper main sequence reflect very likely the need to use a larger, solar-like value for the mixing length parameter,
it is premature, at this stage, to examine in detail the origin of the
discrepancy at faint magnitude, given the still large observational error bars  which range from $\pm 0.04$ to $\pm 0.17$ mag in $M_{606}$ and
$M_{814}$ (Gilmore, priv. com.). If the offset is confirmed by future observations, it reflects very likely the onset of the shortcoming due to the treatment of TiO, which becomes obvious for solar metallicity (see below).

\begin{figure}
\epsfxsize=88mm
\epsfysize=95mm
\epsfbox{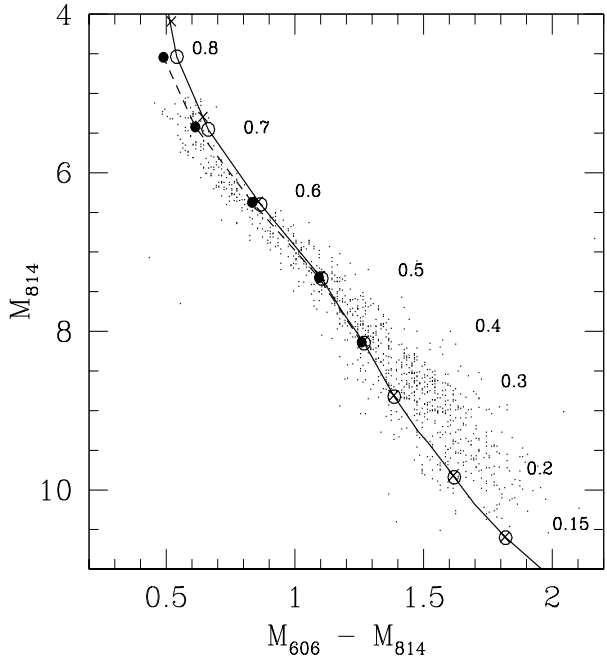}
\caption[ ]{ CMD for 47 Tuc. The data are from Santiago et al. (1996).
The models correspond to [M/H]=-0.5 and different ages and mixing
 length parameters. Solid line (open circles): t=10 Gyrs and $\alpha$=1; dashed line (filled circles): t=10 Gyrs and $\alpha$=2; 
crosses: t=14 Gyrs and $\alpha$=1. Below 0.6 $\msol$, the various symbols are undistinguishable and reflect the negligible effect of age and $\alpha$ for such masses. The masses indicated correspond to the open circles on the solid curve.}
\end{figure}

{$\bullet $} {\it NGC2420 and NGC2477} (Figure 5) : Using the HST
WFPC2 camera, Von Hippel et al. (1996) have obtained deep V- and I-band photometry for the two open clusters NGC 2420 and NGC 2477, with
near-solar metallicity. The instrumental data were kindly provided by T. Von Hippel and transformed into the standard Johnson-Cousins system using the calibration of Holtzmann et al. (1995). NGC2420 is rather metal-poor for an open cluster, with [Fe/H]
$\sim$ -0.45 and NGC2477 is more metal-rich with [Fe/H] $\sim$ 0.
Comparison of theoretical isochrones with such data offers an excellent opportunity to test the present models between [M/H]=-0.5 and 0. We adopt the distance moduli and reddening corrections quoted by Von Hippel et al. (1996), i.e. (m-M)$_0$ = 11.95 and E(B-V) = 0.05 for NGC2420, which corresponds to extinction corrections A$_V$ = 0.155 and
A$_I$ = 0.09, and (m-M)$_0$ = 10.6 and the canonical
value E(B-V) = 0.33 for NGC2477, which gives A$_V$ = 1.01 and A$_I$ = 0.60. Since NGC2477 is known to be differentially reddened, the comparison with theoretical tracks is a delicate task. However, Von Hippel et al. (1996) have examined the effects
of variable reddening E(B-V), from 0.2 to 0.4, and compared their sequence
with the solar metallicity main sequence stars of
Monet et al. (1992), for which trigonometric parallaxes are available. 
The good agreement between both sets of data shows that the
combination of their adopted distance modulus, photometric transformations and reddening correction is reasonable. 

Based on isochrones fitting, the age inferred for NGC2420 ranges from 2 to 4 Gyrs and for NGC2477 from 0.6 to 1.5 Gyrs (Von Hippel et al. 1996, and references therein).  The observed stars, for both clusters, lie in the
range $8\le M_V\le 14$, which corresponds to $0.13 \msol \simle m \simle 0.65 \msol$. Stars in this mass-range have already settled on the main sequence so that age effects are completely negligible from 0.6 to 5 Gyrs. The comparison with observations in Figure 5 is shown for t=1 Gyr isochrones.
As seen, for M$_V \simle $ 10 ($m \simgr 0.5 \msol$)
the data lie well between the two different metallicity sequences,
confirming the agreement obtained for 47 Tuc. Below m=0.5 $\msol$, however, the isochrones are significantly too blue 
and depart by almost 0.5 mag from the NGC2477 observed sequence below $\sim 0.15\,\msol$. 
The reason for such a discrepancy will be examined below.

\begin{figure}
\epsfxsize=88mm
\epsfysize=95mm
\epsfbox{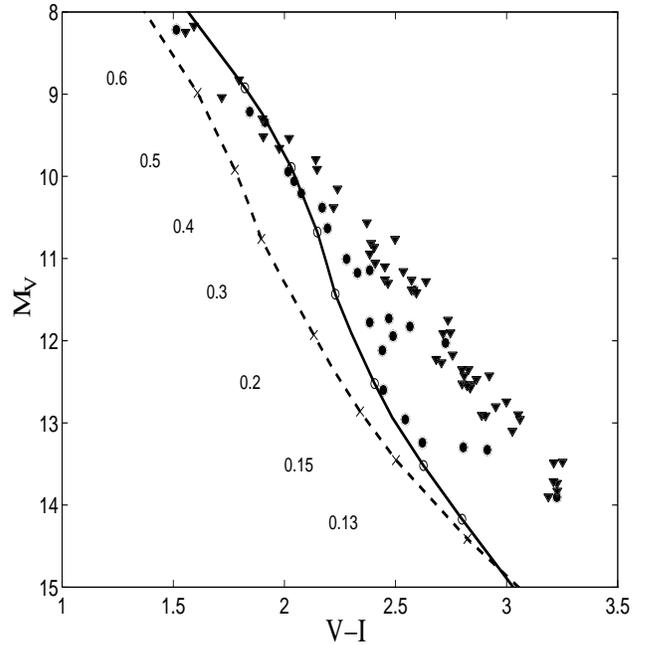}
\caption[ ]{ CMD for NGC2420 (filled circles) and NGC2477 (filled triangles). The data are from Von Hippel et al. (1996), with their
distance modulus and reddening corrections (see text).
The isochrones correspond to [M/H]=-0.5 (dashed line, x) and
[M/H]=0 (solid line, o) for an age t=1 Gyr and a mixing length
parameter $\alpha$ = 1. The masses indicated correspond to the {\it open
circles on the solid curve}. The crosses ([M/H]=-0.5) correspond to lower
masses (see Table 2).
}
\end{figure}
 
\subsection{ Disk field stars}

{$\bullet $} {$\mv$-(V-I) CMD} (Figure 6) : 

Figure 6 displays the observed local sample of (thin)-disk stars of Monet et al. (1992) (full squares) and Dahn et al. (1995) (dots) for which trigonometric parallaxes
have been determined. The sequence of NGC2477 is also shown (full triangles) and is consistent with the disk population,
as mentioned in \S 4.1. 
The theoretical isochrones correspond to
an age t=1 Gyr, for which all stars with $m \ge 0.08 \msol$ have
settled on the main sequence so that age effects for t$>$1 Gyr are negligible.
Figure 6 displays the present models for [M/H]=-0.5 (dashed line) and [M/H]=0 
(solid line), the sequence obtained with the "Base" atmosphere models
(dash-dot line) and the BP95 atmosphere models for [M/H]=0 (hatched line). The discrepancy mentioned 
previously below 0.5 $\msol$ (M$_V \sim 10$; $\te \sim 3600$ K) is obvious down to the bottom of the main sequence, M$_V \sim 20$. This clearly illustrates a real shortcoming in the present models.  The previous "Base" models yield a better agreement
down to M$_V \sim 14$ and are off by $\sim 0.3$ mag only beyond this limit, as already mentioned in BCAH95. We stress, however, that this better agreement is fortuitous and stems from the overestimated opacity in the V-band due to the inaccurate Straight Mean approximation. The models based on BP95 exhibit the same behaviour as the "NextGen" models. 

\bigskip
{$\bullet $} {\it Possible missing source of opacity in the optical:}

Analyzing the possible shortcomings in the models, we first note that the departure between models and observations appears at M$_V \sim 10$, which corresponds to $\te \sim 3600 K$.
This temperature has been pointed out by Leggett et al. (1996)
who compare observed spectra of red dwarfs with the "NextGen" synthetic
spectra. These authors note that a large discrepancy appears for wavelengths shorter than 0.7 $\mu$m below this temperature.
Since TiO dominates the energy distribution of dM stars in the range 0.6-1.1 $\mu$m, the discrepancy may be
due to shortcomings in presently available TiO line lists. VO is
also an important absorber, but only for spectral types later than dM5, which corresponds to m = $0.1 \msol$ and $M_V \sim 16$ ($\te \sim$ 2800 K, cf. Baraffe and Chabrier 1996).
Moreover, comparison between
synthetic and observed spectra of late type M-dwarfs shows an overestimate of
VO absorption features in the R-band of synthetic spectra, rather that an underestimate
required to explain too blue (V-I) colors (Leggett et al. 1996). Therefore, 
uncertainties in the treatment of VO should not be responsible for the departure observed at $M_V \sim 10$ and below.

Alvarez \& Plez (1998) recently analyzed near-infrared photometry of M-giants with an improved version of the Plez et al. (1992) atmosphere models. Although TiO and VO molecular data have been updated, the models still show a discrepancy with observations below $\te \sim 3100K$, as was already the case with the BP95 models, based on the work by Plez et al. (1992) (see hatched line 
in Fig. 6). Alvarez \& Plez (1998)
interpret this departure as an indication for a missing source of opacity
around (shortward of) 1 $\mu$m.
An underestimate of the opacity in this region could be responsible for
 an overestimate of the V-flux by a few tenths of a magnitude, yielding too blue (V-I) colors for a given mass. Even though our mass-$\mv$ relationship agrees well
with the empirical fit of HMC93, such an uncertainty remains within the observational error bars (cf. Fig. 3a).

In order to estimate the effect of a missing opacity and overestimated V-flux
on the synthetic spectra and atmosphere profiles, we have arbitrarily increased
by a factor 5 the total opacity coefficient in the spectral region covering the V-band
($\kappa_\lambda^\prime=5\times \kappa_\lambda$, with $\lambda=$ 0.47 - 0.7 $\mu$m) in order to obtain a fainter V-flux.
The test atmosphere models are performed for
[M/H]=0, $\te$ = 3400 K  and $\te$ = 2800 K
with log g = 5, which roughly correspond to $\mv \sim 11.4$, m $\sim 0.3 \msol$ and
$\mv \sim 16$, m $\sim 0.1 \msol$, respectively,
on the solar metallicty isochrone of Fig. 6. 
The main effect is a redder (V-I) color by 0.6 mag and a fainter $\mv$ by 0.5 mag
for 0.3 $\msol$ and a redder (V-I) by 1.1 mag and fainter $\mv$ by 0.8 mag
for 0.1 $\msol$. Interestingly enough the same increase of opacity translates
into a larger effect at cooler temperature (0.1 $\msol$), as needed.
We stress also that
the atmosphere profiles are hardly affected, suggesting that such an increase of
the opacity in the V-band will not modify the agreement
reached by the present models in the near infrared passbands (see below).
Indeed, the $J,H,K$ bands remains essentially unaffected whereas the I-flux is slightly
increased (0.3 mag at most). Although these calculations are by no means a proof,
they show that a moderate (less than a factor 5) increase of the V-band
opacity would solve the discrepancy between theory and observation for solar metallicity
in the optical below 0.4 $\msol$ without affecting significantly the IR colors.
Note that this (clearly overestimated) increase of the $\mv$-magnitude
yields a mass-$\mv$ relationship which remains within the observational error bars (Fig. 3a),
contrarily to the one obtained with the previous generation of "Base" atmosphere
models.
This simple test only suggests that a missing opacity
in the V-band could improve the models, but a systematic analysis of all
possible sources of uncertainty is required to draw any robust conclusion. 

Regarding possible sources of missing opacity, we note the presence in late type M-dwarf spectra of
a strong absorption feature at $\lambda \sim 0.55 \mu$m due to CaOH 
(see Allard et al. 1997 and references therein), which is not taken into account in the present models. Although we do not expect
this molecule to solve entirely the problem in the V-band, 
the next generation
of atmosphere models should take it into account in order 
to improve the comparison with observed spectra. 
Finally, below
$\te \sim$ 2800 K, grain formation in the outer layers of the atmosphere may affect the spectra, and could be responsible for the discrepancy found at magnitudes fainter than
$M_V \sim 16$ (Tsuji et al. 1996a, Allard 1997a,b). 

In spite of the tremendous improvements performed recently 
in cool-star atmosphere models, the description of molecular absorbers around 1 $\mu$m and $\te < 3700 K$ remains uncertain. 
The solution of this problem, and the treatment of grains, represent the main challenges for the next generation of cool object atmosphere models.

\begin{figure*}
\epsfxsize=130mm
\epsfysize=130mm
\epsfbox{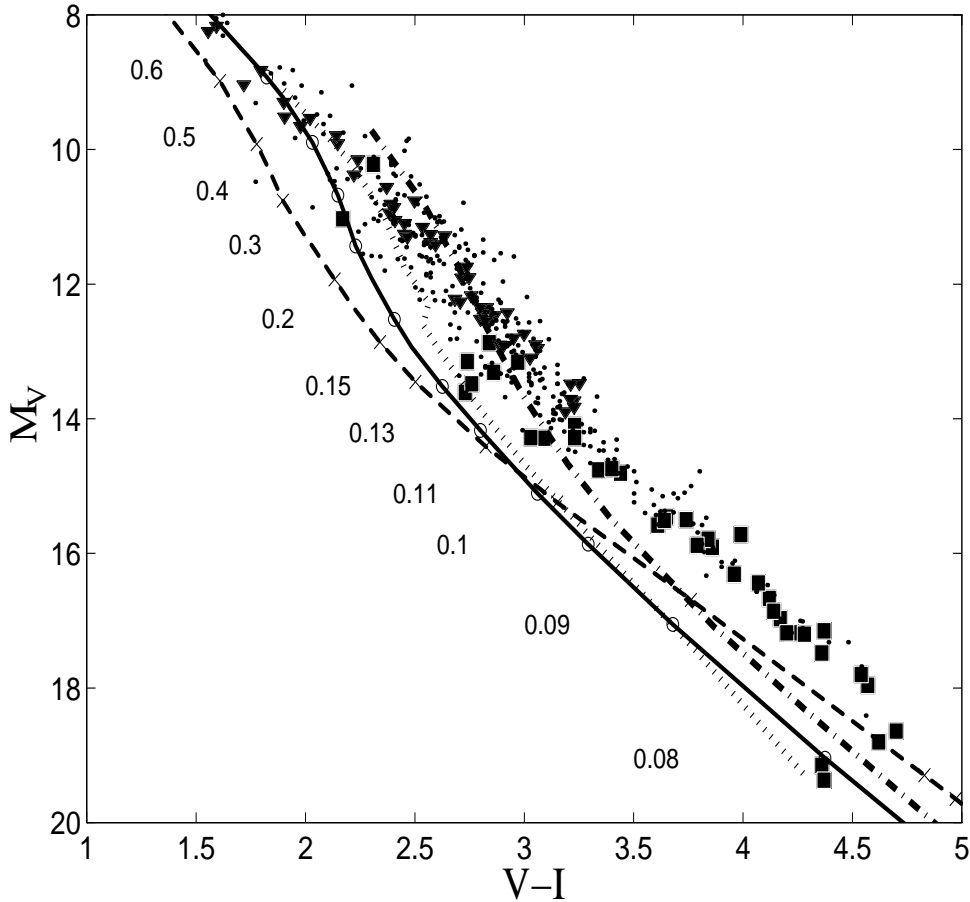}
\caption[ ]{ CMD for disk stars: the data are from Monet et al. (1992) (filled squares) and Dahn et al. (1995) (dots). The sequence of NGC2477 is also shown, with the same distance modulus and reddening as
in Fig. 4 (filled triangles).  Dashed line (x): present models for [M/H]=-0.5.
Solid line (o): present models for [M/H]=0. Dash-dotted line: "Base"
models for [M/H]=0. Hatched line: BP95 models for [M/H]=0.
The isochrones correspond to an age t=1 Gyr. The masses indicated correspond to the open circles on the solid curve.}
\end{figure*}

\bigskip

{$\bullet $} {$\mk$ - (I-K) CMD} (Figure 7) :

The situation in the near-infrared 
is much better and the "NextGen" models
show a real improvement over earlier generations of models. As
shown in \S 3, the mass-$\mk$ relationship agrees well with the observational
empirical relation. Leggett et al. (1996) reach as good an agreement between observed 
and 
synthetic spectra for $\lambda >$ 0.7 $\mu$m and $\te \ge 2700 K$.
The present models reproduce accurately the observed sequences of young clusters
like
the Pleiades and Praesepe (Zapatero Osorio, 1997; Cossburn et al., 1997,
Pinfield et al. 1997). Such an agreement could not be achieved with the stellar models based on the previous "Base" atmosphere models. Regarding field stars, 
Tinney et al. (1993, 1995) carried out a parallax program
on very low mass stars with  available photometry in the I and K passbands.
These objects have all tangential velocities $\simle \, 110$ km.s$^{-1}$ and represent a sample of old disk and young disk
populations. 
Figure 7 shows these data (filled circles), compared to the present models for metallicity ranging from [M/H]= -1.0 to 0. Pleiades objects (Steele et al. 1993, 1995; Zapatero et al. 1997) are also displayed (filled squares) and compared with the 120
Myrs isochrone of the present solar metallicity models (dashed line), 
a reasonable age for this cluster (Basri et al, 1996).
The lowest masses of the 5 Gyr isochrones correspond to the hydrogen burning minimum mass (i.e.  $\sim$ 
0.075, 0.079, 0.083
 $\msol$ for [M/H]=0, -0.5, -1.0, respectively). The 120 Myrs isochrones extend into the brown dwarf regime.

\begin{figure}
\epsfxsize=95mm
\epsfysize=110mm
\epsfbox{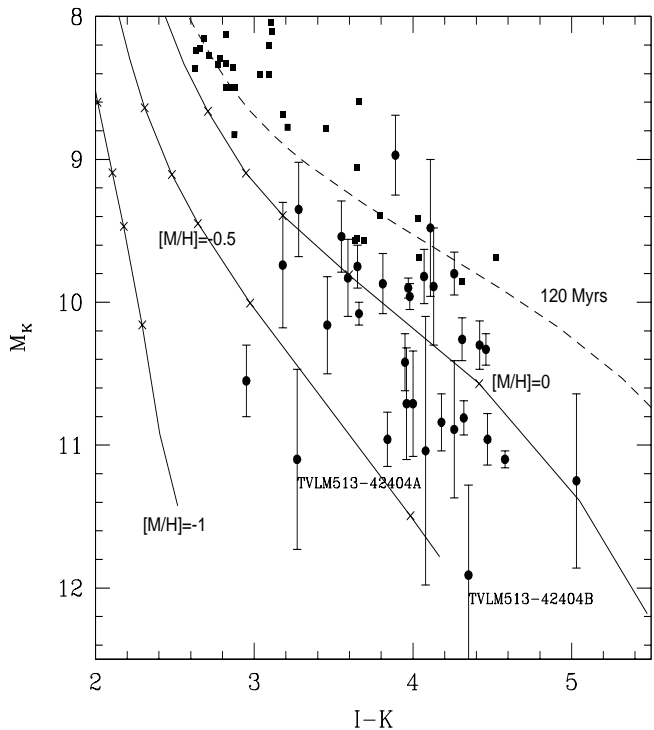}
\caption[ ]{$\mk$-(I-K) CMD: the field stars (full circles) are from Tinney et al. (1993; 1995). The
Pleiades objects (full squares) are from Steele et al. (1993, 1995) and 
Zapatero et al. (1997).
Solid line : present models for t=5 Gyrs and [M/H]=-1, -0.5, 0 from
left to right, as indicated. Dashed line: present models for t=120 Myrs and [M/H]=0.
The crosses on the solid curves correspond to
the following masses : 0.08 (except for [M/H]=-1), 0.09, 0.1, 0.11 and
0.13 $\msol$.}
\end{figure}

As already mentioned and shown in Fig. 7, the bulk of the Pleiades objects, at $\mk \sim 8 - 8.5$ (i.e., $\te \sim 3000K$) agrees fairly well with the 120 Myrs isochrone of the present models.
 Even in the brown dwarf regime, i.e. below 
0.075 $\msol$, which corresponds to $\mk \simgr 9$ and $\te \simle 2800 K$ for this age, the agreement is still reasonable, although the formation of grains at such low temperatures  may alter the spectra. 
This is indeed suggested by the first results of the DENIS survey
which revealed several brown dwarf candidates showing extremely red
(J-K) colors (Forveille et al. 1997). Observed values of (J-K) $> 1$ cannot be reproduced by
the current grainless atmosphere models (cf. Baraffe and Chabrier 1997 and Fig. 8) 
but seem to be in agreement
with preliminary calculations including grain formation 
(Tsuji et al. 1996a; Allard, 1997a,b). 
Although not shown in Figure 7 for sake of clarity, stellar models based on the previous "Base" atmosphere models were too red by $\sim$ 0.3 mag.
 
The Tinney et al. (1993; 1995) sample is particularly interesting, since it covers  
the very bottom of the main sequence, with objects fainter than 
$\mk \sim 8.5$, which corresponds to a solar metallicity 
main sequence star of $\sim 0.13 \msol$ and $\te \simle 3000K$.
Interestingly enough, the data are well distributed between the 
[M/H]=-0.5 sequence and the 120 Myrs [M/H]=0 isochrone, as expected for a mixed old disk/young disk stellar population. This comparison adds credibility to the present models down to the bottom of the main sequence.
Special attention is paid to the 
binary system TVLM513-42404AB discovered by Tinney (1993) which lies at the bottom of the [M/H]=-0.5 sequence, as
indicated in Fig. 7.
A better determination of its parallax would
be highly desirable, although a difficult task as mentioned by Tinney et al. 
(1995). If both members of the system have a common origin (same age and metallicity) and thus can be fitted by a same isochrone, they offer a unique
opportunity to test the predicted shape of the sequence at the brown 
dwarf limit, and to examine the effect of grain formation.  

Finally, we note the powerful diagnostic for metallicity provided 
by $\mk - (I-K)$ CMDs, compared to optical (V-I) colours.  

\bigskip

{$\bullet $} {$\mk$-(J-K) CMD} (Figure 8) : 

Figure 8 displays main sequence ($t>1$ Gyr) and pre-main sequence (0.5 Gyr)
isochrones in the $\mk$ vs $J-K$ CMD for several metallicities ($\mh=0, -0.5, -2.0$)
for the bottom of the MS and down into the BD domain. The dots are the data by
Leggett (1992). We first note the pronounced blue-loop
photometric signature in the IR for objects at the bottom and below the MS, similar to the
one predicted for sub-solar metallicities (see BCAH97). Although this shift of the
flux toward shorter wavelengths is due essentially to the ongoing CIA absorption
of H$_2$ below $\sim 4000$ K ($\sim 0.5\,\msol$) for metal-poor abundances
(Saumon et al., 1994; BCAH97 \S 4.2), it reflects
primarily the formation of methane (CH$_4$) at the expense of CO below $\sim 1800$ K
for solar-like metallicities (Allard et al., 1996; Tsuji et al. 1996b; Marley 
et al. 1996; Burrows et al. 1997). 
Several objects are indicated, the {\it star} VB8, the still undetermined object GD165B (Leggett 1992; Kirkpatrick et al. 1995), the Pleiades BD Teide 1
(Rebolo et al. 1995)  and GL229B (Oppenheimer et al. 1995)
The photometry of GD165B is
severely affected by grain formation and puts it at the very
edge of the stellar/sub-stellar transition (Kirkpatrick et al., 1998).
The agreement between theory and observation is excellent both 
for the MS objects and in the very cool BD regime illustrated by Gl229B (Allard et al., 1996; Burrows et al. 1997). The region in-between, however, is likely
to be strongly affected by the formation of grains in the atmosphere, 
as suggested by the few objects with $\mk \simgr 10$ and $(J-K) > 1$,
although
the general qualitative features will remain the same. Work in this direction
is under progress.

\begin{figure}
\epsfxsize=95mm
\epsfysize=110mm
\epsfbox{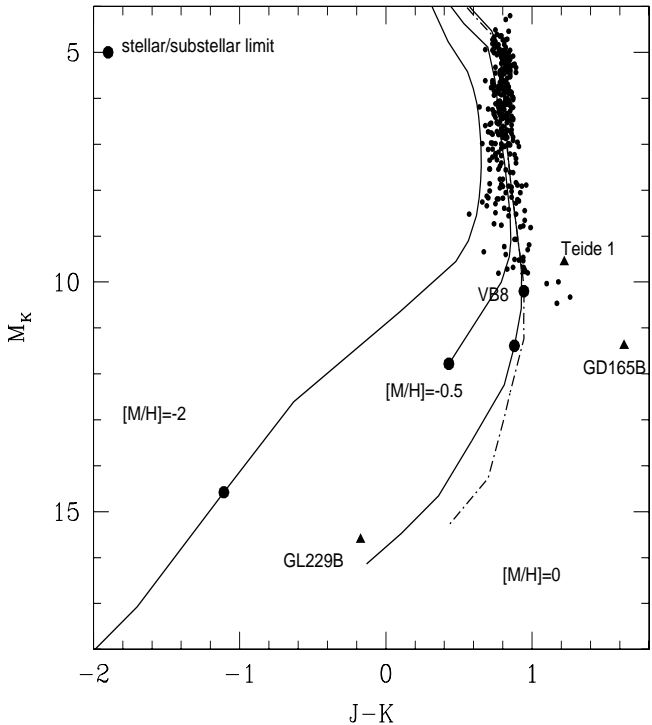}
\caption[ ]{$\mk$-(J-K) CMD: the field stars (dots) are from Leggett (1992).
Several objects are indicated, VB8 and GD165B (Leggett 1992; Kirkpatrick et al. 1995), the Pleiades BD Teide 1
(Rebolo et al. 1995)  and GL229B (Oppenheimer et al. 1995).
Solid line : present models for t=10 Gyrs and [M/H]=-2 , t=6 Gyrs and
[M/H]=-0.5, t=5 Gyrs and [M/H]=0 from
left to right, as indicated. Dash-dotted line: present models for t=0.5 Gyrs and [M/H]=0.
The full circles on the curves correspond to the stellar/substellar
transition: 0.083 $\msol$ for [M/H]=-2, 0.079 $\msol$ for [M/H]=-0.5 and
0.075 $\msol$ for [M/H]=0.
}
\end{figure}

\section{Conclusion}

We have presented solar-type metallicity evolutionary models from 1 $\msol$ down to the hydrogen
burning minimum mass. These models include the most recent interior physics and non-grey atmosphere models and rely on fully consistent calculations between the stellar interior and the atmosphere, with {\it not
a single} adjustable parameter. Any discrepancy between theory and observation thus reflects remaining limitations in the physics entering the theory and not internal inconsistency in the models. In order to carefully examine these limitations, special attention has been paid to the comparison with observed mass-magnitude relationships and color-magnitude diagrams.  

Regarding the optical spectral region, the theoretical mass-$\mv$ relationship
is in excellent agreement with observational data of Henry and McCarthy (1993)
all the way down to the bottom of the main sequence. However, 
the analysis of the $\mv$-$(V-I)$ CMDs highlights the limitation of
the present models for colors redward of (V-I) $\sim 2$, i.e. $\te \simle 3600$ K ($m\simle 0.4\,\msol$), 
predicting (V-I) colors substantially bluer
than the observed ones. This suggests a possible underestimate of 
opacity below 1 $\mu$m. We have tested such an hypothesis by increasing arbitrarily the opacity over the V-band in a couple of
test atmosphere models and obtained
the required effect i.e redder (V-I) color {\it and} unaffected near infrared
fluxes and atmosphere profiles.
Metal-poor models, which 
are less sensitive to metallic molecular absorbers, 
do not suffer from this shortcoming, as illustrated by the remarkable agreement obtained for globular star clusters with [M/H] $\le -1$ (BCAH97). 
The discrepancy begins to appear 
at [M/H]$\sim$ -0.5, as suggested by the comparison with 47 Tuc, although
the
observational error bars remain large, and becomes obvious for solar metallicity. 

This shortcoming is observed also in the
comparison between synthetic and observed M-dwarf (Leggett et al.,
1996) and cool giant (Alvarez \& Plez, 1998) spectra and thus stems
most likely from a still incomplete description of the {\it atmosphere} of
cool objects, rather than from substantial modifications
of their  structural and transport
properties.
 This shortcoming is inherent to all currently available atmosphere models
and represents the next challenge for cool star theorists. 

In the near-infrared, the results are very satisfactory. 
Contrarily to models based on previous generations of atmosphere models,  
the present mass-$\mk$ relationship is in excellent
agreement
with the Henry and Mc Carthy (1993) observational data. The analysis of 
$\mk - (I-K)$ and
$\mk - (J-K)$ CMDs for young open clusters (Zapatero et al., 1997;
Pinfield et al. 1997) and for field disk stars down to the bottom of
the main sequence confirms the significant improvement of the present models
over previous generations.

As for sub-solar metallicities, the photometric signature of the bottom of
the main sequence and of the substellar
domain in the near-IR is a large blueshift due in that case to H$_2$ CIA absorption but also
to the onset of CH$_4$ formation, shifting the peak of the flux to short
wavelengths ($\sim 1\,\mu$m), as observed e.g. in Gl229B (Matthews et al. 1995; Geballe et al. 1996).
The very behaviour of this shift at the high-mass end of the sub-stellar
domain, however, is likely to be affected by grain formation and remains to
be characterized precisely in this region.

At last, we want to stress the following point: comparison between
theory and observation 
in {\it theoretical} $\, \te$-M$_{bol}$ diagrams should be avoided.
Such
comparisons, except for the seldom cases where the exact bolometric magnitude is
determined, are
unreliable since they are based on empirical
color-$\te$ or color-bolometric correction
relations which do not take into account effects of age, gravity or metallicity among
the sample of objects used to derive them.
Although, as stressed in our previous and present calculations, shortcomings are still present in the theory, yielding still slightly
inaccurate bolometric magnitudes, discrepancies arising from comparisons in $\te$-M$_{bol}$ diagrams reflect primarily uncertainties or inconsistencies in the various transformations. Such dubious comparisons, which used to be the only possible ones a few years ago when no {\it synthetic colors} were available, lead in general to incorrect and misleading conclusions both for the theory and the
observations.
Meaningful, consistent comparisons, which avoid external sources of errors, must be done in the various {\it observational} color-magnitude and color-color diagrams.
These latter represent much more stringent constraints for the theory than a
global $\te$-M$_{bol}$ diagram.
Models aimed at describing cool low-mass object structural and thermal
properties must rely
on such a general {\it parameter-free, consistent} theory and must first be proven to
be valid
in {\it every} available passband. With as a holy grail the accurate description
of the physical properties of the star in {\it all} characteristic passbands.

 As shown in this paper, the present
stellar models, although based on updated physics and consistent
interior-atmosphere calculations, still suffer from uncertainties at optical wavelengths,
at least for solar metallicity. This reflects our still uncomplete
knowledge of the complex physics characteristic of cool star-like objects.
It is our aim to solve these shortcomings in a near future
and to derive fully reliable models with completely accurate color- effective temperature
relationships and bolometric correction scales.

\medskip
Tables 1-3 are available by anonymous ftp (including a larger grid in ages):
\par
\hskip 1cm ftp ftp.ens-lyon.fr \par
\hskip 1cm username: anonymous \par
\hskip 1cm ftp $>$ cd /pub/users/CRAL/ibaraffe \par
\hskip 1cm ftp $>$ get BCAH98\_models \par
\hskip 1cm ftp $>$ quit
\bigskip

\begin{acknowledgements} The authors are very grateful to G. Gilmore,
T. Von Hippel, M. Zapatero Osorio, C. Dahn,  C. Tinney and S. Leggett for 
providing their various data. 
This work is funded by grants from  NASA LTSA NAG5-3435 to Wichita State University.
PHH is supported by NAG5-3618 (ATP) and NAG5-3619 (LTSA)
 grants to the University of Georgia.
The calculations presented in this paper were performed at the Cornell
Theory Center (CTC), the San Diego Supercomputer Center (SDSC), and on the
CRAY C94 of the Centre d'Etudes Nucl\'eaires
de Grenoble.
\end{acknowledgements}

\vfill
\eject

\begin{table*}
\caption{Physical properties and absolute magnitudes of low-mass stars for
$[M/H] = 0$, $Y=0.275$ and different ages. The lowest mass corresponds to the hydrogen-burning limit. The mass $m$ is in $M_\odot$, $T_{eff}$ in K. The VRI magnitudes are in the Johnson-Cousins system (Bessell 1990) and the JHK magnitudes in the CIT system (Leggett 1992). Note that the bolometric magnitude corresponds to $M_{bol}(\odot)$ = 4.64.
}
\begin{tabular}{lcccccccccc}
\hline\noalign{\smallskip}
$m$ &age (Gyrs) &$T_{eff}$&$log \,g$ & $M_{bol}$ &$M_V$ &$M_R$ &$M_I$ &$M_J$ &
$M_H$ & $M_K$ \\
\noalign{\smallskip}
\hline\noalign{\smallskip}
 0.075& 0.01&3006.& 4.220&  9.76& 12.60& 11.43&  9.81&  7.83&  7.24&  6.90 \\ 
      & 0.10&2835.& 4.901& 11.72& 15.12& 13.83& 11.96&  9.72&  9.12&  8.79 \\ 
      & 5.00&2003.& 5.393& 14.46& 21.89& 19.24& 16.44& 12.27& 11.56& 11.39 \\ 
 0.080& 0.01&3025.& 4.220&  9.67& 12.46& 11.31&  9.69&  7.74&  7.15&  6.82 \\ 
      & 0.10&2876.& 4.903& 11.59& 14.84& 13.58& 11.76&  9.61&  9.01&  8.69 \\ 
      & 5.00&2313.& 5.350& 13.66& 19.53& 17.54& 14.98& 11.49& 10.82& 10.57 \\ 
 0.090& 0.01&3059.& 4.220&  9.49& 12.21& 11.07&  9.48&  7.57&  6.98&  6.65 \\ 
      & 0.10&2946.& 4.910& 11.38& 14.39& 13.17& 11.44&  9.42&  8.82&  8.51 \\ 
      & 5.00&2641.& 5.292& 12.81& 17.09& 15.56& 13.40& 10.74& 10.10&  9.81 \\ 
 0.100& 0.01&3090.& 4.218&  9.33& 11.97& 10.85&  9.29&  7.42&  6.82&  6.50 \\ 
      & 0.10&3006.& 4.910& 11.17& 13.99& 12.81& 11.15&  9.24&  8.65&  8.34 \\ 
      & 5.00&2812.& 5.251& 12.32& 15.86& 14.50& 12.57& 10.31&  9.69&  9.39 \\ 
 0.110& 0.01&3112.& 4.219&  9.19& 11.79& 10.68&  9.13&  7.29&  6.69&  6.38 \\ 
      & 0.10&3051.& 4.914& 11.02& 13.71& 12.56& 10.94&  9.10&  8.51&  8.20 \\ 
      & 5.00&2921.& 5.219& 11.97& 15.10& 13.84& 12.04& 10.00&  9.40&  9.10 \\ 
 0.150& 0.01&3186.& 4.210&  8.73& 11.19& 10.11&  8.61&  6.85&  6.25&  5.95 \\ 
      & 0.10&3199.& 4.909& 10.46& 12.76& 11.69& 10.23&  8.60&  8.01&  7.72 \\ 
      & 5.00&3151.& 5.133& 11.09& 13.51& 12.41& 10.89&  9.21&  8.62&  8.33 \\ 
 0.200& 0.01&3251.& 4.198&  8.30& 10.65&  9.61&  8.14&  6.44&  5.83&  5.54 \\ 
      & 0.10&3299.& 4.896&  9.98& 12.07& 11.06&  9.68&  8.15&  7.57&  7.29 \\ 
      & 5.00&3292.& 5.066& 10.42& 12.51& 11.48& 10.10&  8.58&  8.01&  7.73 \\ 
 0.300& 0.01&3345.& 4.192&  7.72&  9.93&  8.93&  7.50&  5.88&  5.26&  5.00 \\ 
      & 0.10&3424.& 4.873&  9.32& 11.20& 10.24&  8.95&  7.52&  6.95&  6.69 \\ 
      & 5.00&3436.& 4.966&  9.54& 11.38& 10.42&  9.15&  7.75&  7.18&  6.92 \\ 
 0.350& 0.01&3396.& 4.191&  7.49&  9.62&  8.65&  7.24&  5.66&  5.03&  4.78 \\ 
      & 0.10&3471.& 4.857&  9.06& 10.86&  9.92&  8.66&  7.27&  6.70&  6.45 \\ 
      & 5.00&3475.& 4.929&  9.23& 11.02& 10.07&  8.82&  7.45&  6.88&  6.63 \\ 
 0.400& 0.01&3451.& 4.189&  7.27&  9.34&  8.38&  7.00&  5.45&  4.81&  4.57 \\ 
      & 0.10&3525.& 4.835&  8.79& 10.52&  9.60&  8.37&  7.02&  6.45&  6.20 \\ 
      & 5.00&3522.& 4.888&  8.93& 10.65&  9.72&  8.50&  7.16&  6.59&  6.34 \\ 
 0.500& 0.01&3555.& 4.190&  6.90&  8.84&  7.92&  6.59&  5.11&  4.44&  4.23 \\
      & 0.10&3658.& 4.765&  8.21&  9.77&  8.89&  7.74&  6.48&  5.89&  5.68 \\ 
      & 5.00&3649.& 4.797&  8.30&  9.86&  8.98&  7.83&  6.57&  5.99&  5.77 \\ 
 0.600& 0.01&3645.& 4.194&  6.60&  8.42&  7.55&  6.26&  4.83&  4.15&  3.97 \\ 
      & 0.10&3987.& 4.684&  7.44&  8.61&  7.80&  6.89&  5.80&  5.16&  5.02 \\ 
      & 5.00&3893.& 4.701&  7.58&  8.87&  8.04&  7.05&  5.92&  5.29&  5.13 \\ 
 0.700& 0.01&3719.& 4.193&  6.34&  8.06&  7.22&  5.97&  4.59&  3.90&  3.73 \\ 
      & 0.10&4246.& 4.652&  6.92&  7.80&  7.04&  6.31&  5.36&  4.74&  4.64 \\ 
      & 5.00&4239.& 4.618&  6.84&  7.73&  6.97&  6.24&  5.28&  4.66&  4.56 \\ 
 0.800& 0.01&3795.& 4.192&  6.11&  7.71&  6.89&  5.71&  4.38&  3.68&  3.53 \\ 
      & 0.10&4603.& 4.591&  6.27&  6.82&  6.14&  5.62&  4.84&  4.30&  4.23 \\ 
      & 5.00&4654.& 4.546&  6.11&  6.62&  5.96&  5.46&  4.70&  4.18&  4.11 \\ 
 0.900& 0.01&3882.& 4.191&  5.88&  7.35&  6.56&  5.44&  4.18&  3.48&  3.35 \\ 
      & 0.10&4950.& 4.523&  5.66&  6.00&  5.43&  4.99&  4.34&  3.90&  3.84 \\ 
      & 5.00&5043.& 4.456&  5.41&  5.71&  5.17&  4.74&  4.12&  3.70&  3.65 \\
 1.000& 0.01&4011.& 4.187&  5.62&  6.82&  6.04&  5.10&  3.97&  3.29&  3.17 \\ 
      & 0.10&5265.& 4.460&  5.12&  5.33&  4.85&  4.46&  3.90&  3.53&  3.48 \\ 
      & 5.00&5399.& 4.339&  4.71&  4.88&  4.43&  4.06&  3.53&  3.19&  3.15 \\
\hline
\end{tabular}
\end{table*}

\begin{table*}
\caption{Same as in Table 1 for $[M/H] = -0.5$ and $Y=0.25$  }
\begin{tabular}{lcccccccccc}
\hline\noalign{\smallskip}
$m$ &age (Gyrs) &$T_{eff}$ & log \, g & $M_{bol}$ &$M_V$ &$M_R$ &$M_I$ &$M_J$ &
$M_H$ & $M_K$ \\
\noalign{\smallskip}
\hline\noalign{\smallskip}
 0.079& 0.01&3207.& 4.306&  9.64& 11.79& 10.74&  9.33&  7.81&  7.21&  6.92 \\ 
      & 0.10&3015.& 4.967& 11.56& 14.24& 13.03& 11.37&  9.66&  9.05&  8.78 \\ 
      & 5.00&2025.& 5.445& 14.48& 21.76& 18.81& 15.94& 12.21& 11.80& 11.78 \\ 
 0.080& 0.01&3211.& 4.303&  9.61& 11.75& 10.70&  9.30&  7.78&  7.18&  6.89 \\ 
      & 0.10&3023.& 4.968& 11.54& 14.19& 12.99& 11.34&  9.64&  9.03&  8.76 \\ 
      & 5.00&2130.& 5.434& 14.22& 20.97& 18.26& 15.47& 11.98& 11.52& 11.49 \\ 
 0.090& 0.01&3246.& 4.302&  9.43& 11.50& 10.47&  9.11&  7.62&  7.02&  6.73 \\ 
      & 0.10&3102.& 4.972& 11.31& 13.72& 12.58& 11.04&  9.44&  8.83&  8.56 \\ 
      & 5.00&2714.& 5.347& 12.82& 16.77& 15.09& 12.98& 10.80& 10.19& 10.01 \\ 
 0.100& 0.01&3279.& 4.300&  9.27& 11.27& 10.26&  8.92&  7.46&  6.87&  6.59 \\ 
      & 0.10&3160.& 4.976& 11.12& 13.38& 12.29& 10.81&  9.28&  8.67&  8.41 \\ 
      & 5.00&2941.& 5.293& 12.23& 15.24& 13.89& 12.09& 10.29&  9.68&  9.45 \\ 
 0.110& 0.01&3309.& 4.301&  9.13& 11.07& 10.07&  8.76&  7.34&  6.74&  6.46 \\ 
      & 0.10&3211.& 4.975& 10.95& 13.08& 12.01& 10.59&  9.12&  8.52&  8.25 \\ 
      & 5.00&3070.& 5.255& 11.84& 14.40& 13.20& 11.58&  9.96&  9.35&  9.11 \\ 
 0.150& 0.01&3399.& 4.292&  8.66& 10.43&  9.47&  8.24&  6.89&  6.29&  6.03 \\ 
      & 0.10&3353.& 4.972& 10.42& 12.24& 11.25&  9.98&  8.63&  8.05&  7.79 \\ 
      & 5.00&3312.& 5.160& 10.94& 12.85& 11.83& 10.52&  9.14&  8.55&  8.30 \\ 
 0.200& 0.01&3479.& 4.286&  8.23&  9.87&  8.93&  7.78&  6.49&  5.89&  5.64 \\ 
      & 0.10&3467.& 4.960&  9.93& 11.55& 10.60&  9.44&  8.18&  7.61&  7.36 \\ 
      & 5.00&3453.& 5.090& 10.27& 11.91& 10.95&  9.78&  8.52&  7.94&  7.70 \\ 
 0.300& 0.01&3595.& 4.277&  7.62&  9.09&  8.17&  7.13&  5.92&  5.32&  5.09 \\ 
      & 0.10&3640.& 4.933&  9.21& 10.56&  9.65&  8.66&  7.53&  6.97&  6.73 \\ 
      & 5.00&3643.& 4.989&  9.34& 10.69&  9.78&  8.80&  7.66&  7.11&  6.87 \\ 
 0.350& 0.01&3648.& 4.277&  7.39&  8.78&  7.88&  6.88&  5.71&  5.10&  4.88 \\ 
      & 0.10&3703.& 4.910&  8.91& 10.18&  9.28&  8.35&  7.25&  6.69&  6.47 \\ 
      & 5.00&3694.& 4.942&  9.00& 10.27&  9.37&  8.44&  7.33&  6.78&  6.55 \\ 
 0.400& 0.01&3702.& 4.277&  7.18&  8.50&  7.61&  6.66&  5.52&  4.91&  4.70 \\ 
      & 0.10&3777.& 4.877&  8.59&  9.77&  8.88&  8.02&  6.96&  6.41&  6.19 \\ 
      & 5.00&3756.& 4.903&  8.69&  9.89&  9.00&  8.11&  7.04&  6.49&  6.27 \\ 
 0.500& 0.01&3810.& 4.282&  6.83&  8.02&  7.15&  6.29&  5.20&  4.57&  4.39 \\ 
      & 0.10&4033.& 4.775&  7.81&  8.76&  7.92&  7.21&  6.25&  5.67&  5.51 \\ 
      & 5.00&3956.& 4.788&  7.93&  8.93&  8.08&  7.33&  6.35&  5.78&  5.60 \\ 
 0.600& 0.01&3920.& 4.284&  6.51&  7.60&  6.75&  5.96&  4.91&  4.27&  4.11 \\ 
      & 0.10&4357.& 4.727&  7.16&  7.87&  7.13&  6.55&  5.68&  5.08&  4.98 \\ 
      & 5.00&4359.& 4.690&  7.07&  7.77&  7.04&  6.46&  5.58&  4.98&  4.89 \\ 
 0.700& 0.01&4039.& 4.279&  6.20&  7.19&  6.37&  5.64&  4.63&  3.97&  3.84 \\ 
      & 0.10&4804.& 4.660&  6.40&  6.81&  6.23&  5.76&  5.05&  4.55&  4.50 \\ 
      & 5.00&4876.& 4.618&  6.23&  6.60&  6.05&  5.59&  4.91&  4.43&  4.37 \\ 
 0.800& 0.01&4176.& 4.269&  5.89&  6.75&  5.97&  5.31&  4.35&  3.69&  3.59 \\ 
      & 0.10&5217.& 4.589&  5.72&  5.96&  5.49&  5.09&  4.51&  4.11&  4.06 \\ 
      & 5.00&5342.& 4.524&  5.46&  5.66&  5.22&  4.84&  4.29&  3.91&  3.87 \\ 
 0.900& 0.01&4345.& 4.249&  5.54&  6.26&  5.54&  4.95&  4.04&  3.41&  3.32 \\ 
      & 0.10&5620.& 4.523&  5.11&  5.25&  4.86&  4.51&  4.03&  3.71&  3.68 \\ 
      & 5.00&5780.& 4.405&  4.69&  4.82&  4.45&  4.12&  3.68&  3.38&  3.35 \\ 
 1.000& 0.01&4554.& 4.213&  5.13&  5.70&  5.05&  4.52&  3.69&  3.11&  3.04 \\ 
      & 0.10&5980.& 4.453&  4.55&  4.66&  4.31&  4.01&  3.60&  3.33&  3.31 \\ 
      & 5.00&6160.& 4.253&  3.92&  4.01&  3.69&  3.42&  3.04&  2.79&  2.77 \\ 
\hline
\end{tabular}
\end{table*}

\begin{table*}
\caption{Same as in Table 1 for $[M/H] = 0$ and the inputs required to obtain a solar structure at t=4.61 Gyrs (Y=0.282, $\alpha$=1.9, see text) }
\begin{tabular}{lcccccccccc}
\hline\noalign{\smallskip}
$m$ &age (Gyrs) &$T_{eff}$&$log \, g$ & $M_{bol}$ &$M_V$ &$M_R$ &$M_I$ &$M_J$ &
$M_H$ & $M_K$ \\
\noalign{\smallskip}
\hline\noalign{\smallskip}
 0.700& 0.01&3812.& 4.243&  6.36&  7.92&  7.10&  5.94&  4.64&  3.95&  3.80 \\ 
      & 0.10&4393.& 4.684&  6.85&  7.58&  6.84&  6.22&  5.34&  4.75&  4.67 \\ 
      & 5.00&4418.& 4.652&  6.74&  7.45&  6.72&  6.12&  5.25&  4.66&  4.58 \\ 
 0.800& 0.01&3945.& 4.265&  6.12&  7.45&  6.66&  5.64&  4.45&  3.77&  3.64 \\ 
      & 0.10&4819.& 4.638&  6.19&  6.60&  5.99&  5.53&  4.83&  4.35&  4.29 \\ 
      & 5.00&4909.& 4.598&  6.01&  6.37&  5.79&  5.35&  4.67&  4.22&  4.17 \\ 
 0.900& 0.01&4126.& 4.277&  5.83&  6.87&  6.10&  5.27&  4.23&  3.58&  3.47 \\ 
      & 0.10&5234.& 4.588&  5.58&  5.81&  5.31&  4.92&  4.35&  3.97&  3.93 \\ 
      & 5.00&5369.& 4.525&  5.31&  5.50&  5.03&  4.66&  4.13&  3.78&  3.74 \\ 
 1.000& 0.01&4325.& 4.280&  5.52&  6.31&  5.57&  4.91&  4.00&  3.39&  3.30 \\ 
      & 0.10&5596.& 4.536&  5.04&  5.18&  4.76&  4.41&  3.94&  3.63&  3.59 \\ 
      & 5.00&5814.& 4.429&  4.61&  4.71&  4.32&  4.01&  3.58&  3.31&  3.28 \\ 
\hline
\end{tabular}
\end{table*}

\end{document}